\documentstyle[11pt]{article}
\begin{document}

\newcommand{\vm}{\vspace{0.2cm}}
\newcommand{\vl}{\vspace{0.4cm}}

\vspace{0.2cm}
\begin{center}
\begin{Large}
{\bf Renormalization Constant of the Color Gauge Field\\ as a Probe
of Confinement}
\end{Large}
\vskip 10mm { \small {\bf Masud Chaichian}}\\
      {\small   Department of Physics, High Energy Physics Division}\\
           {\small University of Helsinki} \\
{\small and}\\ {\small Helsinki Institute of Physics}\\ {\small
P.O. Box 9, FIN-00014 Helsinki, Finland}\\ {\small and} \\ {
\small {\bf Kazuhiko Nishijima}}\\ {\small Nishina Memorial
Foundation}\\ {\small  2-28-45 Honkomagome, Bunkyo-ku, Tokyo
113-8941, Japan}
\date{}
\end{center}


\vspace{2.0cm}
\begin{center}\begin{minipage}{5in}
\begin{center} {\bf Abstract} \end{center}
\baselineskip 0.3in {\small The mechanism of color confinement as
a consequence of an unbroken non-abelian gauge symmetry and
asymptotic freedom is elucidated and compared with that of other
models based on an analogy with the type II superconductor. It is
demonstrated that a sufficient condition for color confinement is
given by $Z_3^{-1}=0$ where $Z_3$ denotes the renormalization
constant of the color gauge field. It is shown that this condition
is actually satisfied in quantum chromodynamics and that some of
the characteristic features of other models follow from it.}\\

\vskip 10mm
\end{minipage}
\end{center}


\section{Introduction}

It is our consensus that strong interactions are governed by
quantum chromodynamics (QCD) or the gauge theory of quarks and
gluons. These fundamental constituents of hadrons carry the color
quantum number and are considered to be unobservable. This is a
conclusion drawn from our unsuccessful attempts to observe
isolated quarks and is referred to as color confinement -
abbreviated as c.c. - in what follows. It is the subject of this
paper to compare various interpretations of c.c. in an effort to
extract common features from them. For this purpose we review
early attempts to interpret confinement in Section 2. These
theories are formulated in configuration space and are
characterized by the two phase structure or the dual Meissner
effect with a finite penetration depth. In Section 3 we
recapitulate an interpretation of confinement formulated in the
state vector space on the basis of BRS invariance and asymptotic
freedom (AF) skipping details of the proof. In this section we
present two alternative conditions for confinement. In Section 4
we prove that one of these two conditions is a consequence of the
other on the basis of the renormalization group (RG) method. It is
shown that c.c. is realized when the condition $Z_3^{-1}=0$ is
satisfied where $Z_3$ is the renormalization constant of the color
gauge field. It is called the condition for color confinement
abbreviated as CCC hereafter. In Section 5 it is proved that the
CCC is really satisfied in QCD.

Finally in Section 6 we compare the consequences of the CCC with
other interpretations. First, we give an intuitive interpretation
of the CCC in fictitious electrodynamics. Next, we discuss the
connection between the CCC and the linear potential between a
heavy quark and a heavy antiquark pair resulting from Wilson's
area law. Then, we show that the flux of the color gauge field
emerging from color singlet hadrons cannot penetrate into the
confining vacuum leaving no trace of long range forces. This
resembles the dual Meissner effect introduced in some of the other
interpretations.

Two appendices are included. Appendix A is intended to clarify the relationship
between two alternative forms of the conditions for color confinement. Appendix
B gives a derivation of (3.48) in the absence of asymptotic fields due to
infrared singularities. 

\section{Early Attempts to Interpret Confinement}

Because of its profound mysterious nature exhibited in strong interactions
various attempts have been made to understand the mechanism of color
confinement on the basis of classical or semi-classical gauge theories
eventually exploiting topological quantization.

Starting from a classical Lagrangian of the Higgs model, Nielsen
and Olesen [1] identified the hadronic strings with the
Landau-Ginzburg-Abrikosov vortices of quantized magnetic flux in
the superconducting vacuum. Their vortices are either endless or
closed and the energy of the system is minimized for a certain
optimum radius of the vortex, and the total flux is topologically
quantized.

Nambu [2] introduced Dirac's monopoles into this theory and
realized finite vortices by putting monopoles at both ends, and
Dirac's quantization for monopoles matches the flux quantization.
Thus a hadronic string is formed in a superconducting vacuum by
joining a monopole-antimonopole pair by a vortex of a quantized
magnetic flux. The flux cannot spread out and the energy of the
flux is proportional to the length of the string. This pair may be
regarded as a meson and monopoles in the superconducting vacuum
are then confined. An ordinary superconductor is a coherent
superposition of charged objects, but a vacuum which confines
electric, instead of magnetic, monopoles may be a coherent
superposition of magnetic monopoles  as suggested by Mandelstam
[3] and by 't Hooft [4].  Magnetic monopoles do not appear in
continuum quantum electrodynamics (QED) unless they are put by
hand, so that electric charges cannot be confined in QED if this
interpretation should be taken for granted. They can appear in
non-abelian gauge theories, however,  as shown by 't  Hooft [5],
and therefore the Yang-Mills vacuum could be a coherent
superposition of magnetic monopoles, and confinement must be a
property characteristic of non-abelian gauge theories. We shall
come back to this subject later again in Subsections 3.4 and 6.1.

In the quantized version of non-abelian gauge theories, the only
known way of including higher order corrections in a
gauge-invariant manner is to exploit the lattice gauge theory [6],
although Lorentz invariance is recovered only in the limit of the
vanishing lattice constant. Since confinement is a
non-perturbative effect it is important to check it in the lattice
gauge theory. Wilson [6] has formulated confinement in the form of
the area law for the loop correlation function leading to a
confining linear potential between a quark and an antiquark. In
the strong coupling approximation the area law is obeyed even in
QED, but it is a non-trivial problem to check if the strong
coupling regime could be continued to the weak coupling one
without  encountering a phase transition. Susskind and Kogut [7]
analyzed the lattice gauge theory in the strong coupling
approximation and found that the confining strong coupling phase
resembles that of a type II superconductor with electric and
magnetic fields interchanged. We shall come back to the discussion
of the area law later again in the Subsection 6.2.

All such attempts including the recent supersymmetric Seiberg-Witten theory
[8] tend to indicate that confinement is a consequence of the coherent
superposition
of magnetic monopoles in the vacuum state dual to the superconducting one
based on coherent superposition of charged objects such as Cooper pairs.
Quarks and antiquarks are joined together by the electric flux penetrating
into the vacuum or through the normal conducting phase enclosed by the
superconducting phase. In other words, confinement is characterized by the
two-phase structure or the Meissner effect of a finite penetration depth.

In electrodynamics all the physical media are characterized by their
dielectric constant $\epsilon >1$. Suppose that there is a fictitious medium
of dielectric constant [9],
$$
\begin{array}{lll}
\epsilon \ll 1 .
\end{array}
\eqno{(2.1)} $$ \noindent This medium is antiscreening. When a
small charge is placed in this medium, it will crack and develop a
hole surrounding this charge. We have $\epsilon=1$ inside and
$\epsilon<<1$ outside. Because of the antiscreening nature of the
medium the induced charge on the inner surface of the hole is of
the same sign as the originally inserted charge. In order to
reduce the size of the hole the repulsion between the original
charge and the induced one must be overcome, but elimination of
the hole would require infinite energy, so that the hole will not
disappear. The two media, one with $\epsilon=1$ and the other with
$\epsilon <<1$, are considered to correspond to normal conducting
and superconducting states, respectively, and such a two-phase
structure of the vacuum is common to all the models constructed in
analogy with the type II superconductor in the configuration
space.

 In the next section we shall recapitulate the arguments based on the BRS
invariance [10] and asymptotic freedom [11,12] in the state vector space.

\section{Color Confinement as a Renormalization Effect}

In a covariant quantization of gauge fields introduction of
indefinite metric is indispensable. Thus the resulting state
vector space ${\cal V}$ involves unphysical states of indefinite
metric and we have to find a criterion to select physical states
out of ${\cal V}$.  For this purpose we employ the Lorentz
condition in QED as a subsidiary condition, but it is more
involved in non-abelian gauge theories. In what follows we shall
confine ourselves to QCD, and in order to fix the notation we
start from its Lagrangian density in the Pauli metric,

$$
\begin{array}{lll}
{\cal L} = {\cal L}_{inv}+ {\cal L}_{gf}+{\cal L}_{FP},
\end{array}
\eqno {(3.1)} $$ \noindent where
 $$\begin{array}{lll}
\hspace{1.0cm}&     {\cal L}_{inv}= -\frac{1}{4}F_{\mu\nu}\cdot
F_{\mu\nu} - \overline{\psi}(\gamma_{\mu} D_{\mu}+m)\psi
, & \hspace{6.0cm} (3.2a) \\ &\\ \hspace{1.0cm}&   {\cal
L}_{gf}=A_{\mu}\cdot \partial_{\mu} B +\frac{\alpha}{2} B\cdot B ,
& \hspace{6.0cm} (3.2b)\\ &&\\ \hspace{1.0cm}& {\cal L}_{FP}=
i\partial_{\mu} \overline{c} \cdot D_{\mu}c .  &
\hspace{6.0cm}(3.2c)\\
\end{array}$$

\noindent  We have suppressed the color and flavor indices above.
The first term in (3.1) is the gauge-invariant term, the second
one the gauge-fixing term and the last one the Faddeev-Popov ghost
term. In (3.2b) $\alpha$ denotes the gauge parameter and $B$ the
Nakanishi-Lautrup auxiliary field, and in (3.2c) the hermitian
scalar fields $c$ and $\bar{c}$ are anticommuting and are called
Faddeev-Popov (FP) ghost fields.

In what follows we shall introduce the inner and outer products of
two colored objects:
\begin{eqnarray*}
\hspace{1.0cm} S \cdot
T= \sum_{a} S^{a} T^{a}, \hspace{10.3cm}(3.3)
\end{eqnarray*}
\begin{eqnarray*} \hspace{1.0cm} (S \times T)^a= \sum_{b,c}
f_{abc} S^{b} T^{c}, \hspace{9.0cm}(3.4)
\end{eqnarray*}
\noindent where $a$, $b$, $c$ etc. are color indices and $f_{abc}$
the structure constant of the algebra su(3) corresponding to the
color gauge group. Then covariant derivatives are defined by
$$\begin{array}{lll} \hspace{1.0cm}& D_{\mu} \psi = (\partial
_{\mu} -i g T \cdot A_{\mu} ) \psi, \hspace{7.8cm}(3.5)\\
\end{array}$$
$$\begin{array}{lll} \hspace{1.0cm}& D_{\mu} c^a = \partial _{\mu}
c^a + g (A_{\mu} \times c)^a, \hspace{7.5cm}(3.6)\\
\end{array}$$
$$\begin{array}{lll} \hspace{1.0cm}& F_{\mu\nu}^a = \partial
_{\mu} A_{\nu}^a - \partial _{\nu} A_{\mu}^a + g (A_{\mu} \times
A_{\nu})^a. \hspace{6.0cm}(3.7)\\
\end{array}$$

The local gauge invariance is respected by (3.2a) but not by the
other two, (3.2b) and (3.2c), introduced for quantization. It so
happens, however, that the total Lagrangian density is invariant
under new global transformations called the Becchi-Rouet-Stora
(BRS) transformations [10].

\subsection{BRS transformations}

Let us consider an infinitesimal gauge transformation of the gauge and
quark
fields and replace the infinitesimal gauge function either by $c$ or
$\overline{c}$.
They define two kinds of BRS transformations denoted by $\delta$ and
$\overline{\delta}$, respectively.

$$
\begin{array}{lll}
\delta A_{\mu}= D_{\mu}c , & \overline{\delta} A_{\mu}=
D_{\mu}\overline{c} ,\\
\end{array}
\eqno{(3.8)}
$$
$$
\begin{array}{lll}
\delta \psi= ig (c\cdot T) \psi ,& \overline{\delta}  \psi= ig
(\overline{c}\cdot T) \psi .\\
\end{array}
\eqno{(3.9)}
$$
\noindent For the auxiliary fields $B$, $c$ and $\bar{c}$ local
gauge transformations are not even defined, but their BRS transformations
can be introduced by requiring the invariance of the local Lagrangian
density, namely,

$$
\begin{array}{lll}
\delta {\cal L} = \overline{\delta} {\cal L}= 0,
\end{array}
\eqno{(3.10)} $$ \noindent then we find the following
transformations:

$$
\begin{array}{lll}
\delta B = 0, \delta \bar{c} = i B,  \delta c = - \frac{1}{2} g (c \times
c),
\end{array}
\eqno{(3.11a)}
$$
$$
\begin{array}{lll}
\overline{\delta} \bar{B} = 0, \overline{\delta} c = i \bar{B},
\overline{\delta} \bar{c} = - \frac{1}{2} g (\bar{c} \times
\bar{c}),
\end{array}
\eqno{(3.11b)} $$ \noindent where $\bar{B}$ is defined by $$
\begin{array}{lll}
B + \bar{B} - ig (c\times \bar{c}) = 0.
\end{array}
\eqno{(3.12)}
$$

Noether's theorem states that the BRS invariance of the Lagrangian
density leads to conservation laws and we introduce two conserved
BRS charges $Q_B$ and $\overline{Q}_B$ by

$$
\begin{array}{lll}
\delta \phi= i \left[ Q_B,\phi\right]_{\mp} , &  \overline{\delta}
\phi=
i \left[ \overline{Q}_B,\phi\right]_{\mp}    ,
\end{array}
\eqno{(3.13)}
$$
\noindent where we choose the $-(+)$ sign, when the field $\phi$ is of an
even
(odd) power in the ghost fields $c$ and $\overline{c}$.

Maxwell's equations for the gauge field can be expressed in terms of BRS
transformations as

$$
\begin{array}{lll}
\partial_{\mu}F_{\mu\nu} + gJ_{\nu}= i\delta\overline{\delta}
 A_{\nu} ,
\end{array}
\eqno{(3.14)}
$$
 \noindent where $J_{\nu}$ denotes the color current density and $g$ the
gauge coupling constant. This set of equations has been derived from
the Lagrangian density (3.1), but the form of these equations is valid
for a wider class of theories such as supersymmetric and grand unified
theories as long as the original gauge symmetry is respected. The only
deviation from the original theory represented by (3.1) appears when
we try to express the color current density $J_{\nu}$ explicitly in
terms of the elementary component fields.

The r.h.s. of Eq. (3.14) represents a
conserved current and stands for the deviation from Maxwell's classical
equations.

$$
\begin{array}{lll}
\partial_{\mu}(i\delta\overline{\delta}
 A_{\mu}) =0 .
\end{array}
\eqno{(3.15)} $$ \noindent The BRS charges are hermitian and
nilpotent, for example,

$$
\begin{array}{lll}
Q_B^{\dagger}= Q_B , & Q_B^2=0 .\\
\end{array}
\eqno{(3.16)}
$$
\noindent The nilpotency implies introduction of indefinite metric and a
physical state $\vert f\rangle$ is defined by a constraint [14]

$$
\begin{array}{lll}
Q_B\vert f\rangle =0  , &  \vert f\rangle \in {\cal V}  .\\
\end{array}
\eqno{(3.17)}
$$
 \noindent The collection of physical states including the vacuum state
$\vert 0\rangle$ forms the physical subspace of ${\cal V}$ denoted by ${\cal
V}_{phys}$,

$$
\begin{array}{lll}
{\cal V}_{phys} =\{ \vert f\rangle: Q_B\vert f\rangle =0  ,   \vert
f\rangle \in {\cal V}\}  .
\end{array}
\eqno{(3.18)} $$ \noindent It should be mentioned that the S
matrix is BRS invariant,

$$
\begin{array}{lll}
\delta S = i\left[ Q_B, S \right] =0  ,
\end{array}
\eqno{(3.19)} $$ \noindent so that the  physical subspace ${\cal
V}_{phys}$  is an invariant subspace of the S matrix.

Furthermore, we introduce a subspace of ${\cal V}$ called the daughter
subspace
${\cal V}_d$ defined by

$$
\begin{array}{lll}
{\cal V}_{d} =\{ \vert f\rangle: \vert f\rangle=Q_B\vert g\rangle
,   \vert g\rangle \in {\cal V}\}  .\\
 \end{array}
\eqno{(3.20)} $$ \noindent Because of the nilpotency of $Q_B$,
${\cal V}_d$ is a subspace of ${\cal V}_{phys}$,

$$
\begin{array}{lll}
{\cal V}_{d} \subset {\cal V}_{phys},
 \end{array}
\eqno{(3.21)} $$ \noindent and we introduce the Hilbert space
${\cal H}$ by

$$
\begin{array}{lll}
{\cal H}= {\cal V}_{phys}/{\cal V}_{d}   ,
 \end{array}
\eqno{(3.22)} $$
\noindent which may be called the BRS cohomology
[15,16,17].

\subsection{Relation to QED}

QED is the oldest example of gauge theories so that the above
formulation should be applicable to it. We recognize that the
subsidiary condition (3.17) to select physical states looks
completely different from the Lorentz condition,
$$
\begin{array}{lll} B^{(+)} (x) \vert f \rangle = 0,
\end{array}
\eqno{(3.23)} $$

\noindent so that we shall clarify the relationship between them.

In QED or in an abelian gauge theory the auxiliary fields $B$, $c$
and $\bar{c}$ are free and massless, namely, $$
\begin{array}{lll}
\Box B(x) = \Box c(x) = \Box \bar{c} (x) = 0.
\end{array}
\eqno{(3.24)} $$ \noindent Furthermore, in the conventional
treatment of QED the ghost fields do not participate in the game
so that we introduce constrained physical states [18] in terms of
the positive frequency parts of ghost fields by

$$
\begin{array}{lll}
Q_B \vert f^{'} \rangle = 0,
\end{array}
\eqno{(3.17)} $$

$$
\begin{array}{lll}
c^{(+)}(x) \vert f^{'} \rangle =
\bar{c} ^{(+)} (x) \vert f^{'} \rangle =0,
\end{array}
\eqno{(3.25)} $$

\noindent and the corresponding subspace ${\cal V}_{phys}^{'}$.
Consistency among them requires  the following condition:

$$
\begin{array}{lll}
\{ Q_{B}, \bar{c}^{(+)} (x) \} \vert f^{'} \rangle = B^{(+)} (x)
\vert f^{'} \rangle = 0,
\end{array}
\eqno{(3.26)} $$ \noindent which is precisely the Lorentz
condition.

On the other hand, when we have the Lorentz condition and (3.25)
the condition (3.17) follows automatically from the structure of
$Q_B$ in QED, namely,

$$
\begin{array}{lll}
Q_B = \int d^3x [B^{(-)} (x) \dot{c} ^{(+)} (x) - \dot{B} ^{(-)}
(x) c^{(+)} (x) + \dot{c} ^{(-)} (x) B^{(+)} (x) - c^{(-)} (x)
\dot{B} ^{(+)} (x)].
\end{array}
\eqno{(3.27)} $$ \noindent Thus we realize that the two conditions
(3.17) and (3.23) are equivalent under the constraints (3.25).
When we define the constrained daughter space ${\cal V}_d^{'}$ as
a subspace of ${\cal V}_d$ constrained by (3.25), we have

$$
\begin{array}{lll}
{\cal V}_{phys}^{'}/{\cal V}_{d}^{'} = {\cal V}_{phys}/{\cal
V}_{d} = {\cal H}.
\end{array}
\eqno{(3.28)} $$ \noindent This argument indicates that the
condition (3.17) is not alien to QED [19].

\subsection{Interpretation of color confinement}

 When single-quark states and
single-gluon states are not physical, they are unobservable and
hence confined. Thus the problem of color confinement is settled
if we could prove

$$
\begin{array}{lll}
Q_B\vert quark\rangle \neq 0  , &  Q_B\vert gluon\rangle \neq 0  .\\
\end{array}
 \eqno{(3.29)}
$$

\noindent This is a definition of c.c. given in terms of
unobservable quantities so that we shall present an alternative
definition of c.c. in terms of observable quantities in Section 5.

In order to study the condition on which the relations (3.29) are
satisfied we start from an identity:

$$
\begin{array}{lll}
\langle A_{\mu}^a(x),B^b(y) \rangle &\equiv&
   \langle 0 \vert T\left[A_{\mu}^a(x),  B^b(y)
\right]\vert 0\rangle\\
&
=& -\delta_{ab} \partial_{\mu}D_F(x-y) ,\\\end{array} \eqno{(3.30)}
$$

\noindent where $a$ and $b$ denote color indices and $D_F$ is the free
massless propagator. This identity implies that both
$A_{\mu}$ and $B$ generate a massless spin zero particle as applied to the
vacuum state. Thus their asymptotic fields may be expressed as

$$
\begin{array}{lll}
 A_{\mu}^a(x)^{in}= \alpha_{\mu}^a(x) + \partial_{\mu}\chi^a(x) ,
& B^b(y)^{in}=\beta^b(y) , \\
\end{array}
\eqno{(3.31)} $$

\noindent where $\alpha_{\mu}$ denotes the incoming gluon field
and $\chi$ and $\beta$ are the incoming fields of the massless
spin zero particle. Here, we have assumed the validity of the
Lehmann-Symanzik-Zimmermann (LSZ) asymptotic condition [20] or its
suitable modification so that we can relate a field operator to a
particle state through its asymptotic field. 
Since QCD is
infested with infrared singularities an infrared cut-off is introduced
to validate the asymptotic condition, and only after confinement of
colored particles it can be lifted safely for the system of hadrons.
This is related to the Meissner-like effect discussed in Subsection
6.3.

The asymptotic fields introduced above satisfy the relations,

$$\begin{array}{lll}
\partial_{\mu}\alpha_{\mu}^{a} = 0  , & \Box^{2} \alpha_{\mu}^a=0 ,
 \end{array}
\eqno{(3.32)} $$
$$\begin{array}{lll} \Box \chi^a = \alpha \cdot
\beta^a  , & \Box \beta^a=0 ,
\end{array}
\eqno{(3.33)} $$

\noindent and

$$
\begin{array}{lll}
 \langle \chi^a(x), \beta^b(y)\rangle= -\delta_{ab} D_F(x-y) ,
\end{array}
\eqno{(3.34)} $$ 

$$
\begin{array}{lll}
\langle \beta^a(x), \beta^b(y)\rangle=    0 ,
\end{array}
\eqno{(3.35)} $$

\noindent Now we shall study the implications of c.c. in the
properties of the asymptotic fields. Assume that
$\alpha_{\mu}^{a}$ is BRS invariant, namely, $$
\begin{array}{lll}
\delta \alpha _{\mu}^a=0,
\end{array}
\eqno{(3.36)} $$ \noindent then $$
\begin{array}{lll}
Q_B \vert gluon \rangle = Q_B \alpha_{\mu}^{a} \vert 0 \rangle =
(-i) \delta \alpha_{\mu}^{a} \vert 0 \rangle = 0,
\end{array}
$$ \noindent so that a single-gluon state is physical and hence
observable. Therefore, c.c. implies $$
\begin{array}{lll}
\delta \alpha _{\mu}^a \neq 0.
\end{array}
\eqno{(3.37)} $$ \noindent Now $$
\begin{array}{lll}
\delta A_{\mu}^{a,in}&=&\delta \alpha_{\mu}^{a} + \partial _{\mu}
(\delta \chi ^{a})\\ &=& \partial _{\mu} c^{a,in} + g (A_{\mu}
\times c )^{a,in}.
\end{array}
\eqno{(3.38)} $$ \noindent When the second bilinear term on the
r.h.s. of Eq. (3.38) is absent, there is no asymptotic field of
unit spin in this expression and we are led to Eq. (3.36). This is
really the case in perturbation theory. Therefore, a necessary
condition for gluon confinement is the non-vanishing of $(A_{\mu}
\times c )^{in}$. This means that there must be a bound state
between a gluon and a FP ghost, and then and only then we have
(3.37) and gluons are confined. Similarly, quarks are confined
when a quark and a FP ghost form a bound state. Thus the problem
of confinement reduces to that of bound states.

One of the present authors (KN) and Okada studied the bound state
problem by making use of the Bethe-Salpeter equations in the
ladder approximation and recognized that quarks and probably also
gluons are confined when two FP ghosts form a bound state [21].
The statement that c.c. is a consequence of the formation of the
dighost bound state was plausible but not conclusive because of
the approximate nature of the above treatment. This condition was
further refined and reappeared later in a simpler form, namely,
the condition (3.49) to be introduced in the next subsection. It
is not difficult to show that the formation of dighost bound
states is a consequence of this new condition.
This relationship is clarified in Appendix A.

\subsection{Conditions for color confinement}

 By utilizing the conserved current (3.15) we shall introduce a
 set of Ward-Takahashi identities. First, we define the conserved
 color charge $Q^a$ in terms of the color current density
 $J_{\nu}^{a}$ in Eq. (3.14) by
$$
\begin{array}{lll}
Q^a = \int d^3 x J_{0}^{a} (x).
\end{array}
\eqno{(3.39)} $$ \noindent Then let us consider a colored field
$\Phi^{\alpha}$ belonging to an irreducible representation $R^a$
of the Lie algebra su(3) of the color symmetry, $$
\begin{array}{lll}
\left [\Phi ^{\alpha} (y), Q^a \right] &=&R^a_{\alpha\beta}
\Phi^{\beta} (y),\\ \left [\bar{\Phi} ^{\alpha} (z), Q^a \right]
&=&-\bar{\Phi} ^{\beta}(z) R^a_{\beta\alpha} = -
(R^a)^{T}_{\alpha\beta} \bar{\Phi}^{\beta} (z) ,\\
\end{array}
\eqno{(3.40)} $$ \noindent where $\bar{\Phi}$ is the adjoint of
$\Phi$ and $R^{a}$ is an hermitian matrix.

For the quark field $R^{a}$ is given by $T^{a}$ in Eq. (3.5) or
$\lambda^{a}/2$, where $\lambda^{a}$ denotes Gell-Mann's matrix.
For the color gauge field obeying the adjoint representation we
have $R^{a}_{bc}=-if_{abc}$. Then we have a set of Ward-Takahashi
identities of the form

$$
\begin{array}{lll}
&&\partial _{\mu} \langle \delta \bar{\delta} A_{\mu}^{a} (x),
\Phi^{\alpha} (y), \bar{\Phi}^{\beta} (z) \rangle \\ && =ig \left
[ R_{\alpha\gamma}^{a} \delta ^4 (x-y) \langle \Phi^{\gamma} (y),
\bar{\Phi}^{\beta} (z) \rangle - R_{\delta\beta}^{a} \delta ^4
(x-z) \langle \Phi^{\alpha} (y), \bar{\Phi}^{\delta} (z)
\rangle\right ].
\end{array}
\eqno{(3.41)} $$ \noindent This identity follows from Maxwell's
equations. When $\Phi$ represents a color singlet field we have
$R=0$ and the r.h.s. of Eq. (3.41) vanishes identically.

In order to derive a condition for color confinement we need some
preparations. When an operator $M$ is the BRS transform of another
operator $N$, namely,

$$
\begin{array}{lll}
M = \delta N = i \left [ Q_B,N \right]_{\mp},
\end{array}
\eqno{(3.42)} $$ \noindent $M$ is called an exact operator. Thus
its matrix element between two physical states $\vert \alpha
\rangle$ and $\vert \beta \rangle$ vanishes identically because of
the definition of the physical states (3.17).

$$
\begin{array}{lll}
\langle \beta \vert M \vert \alpha \rangle = 0.
\end{array}
\eqno{(3.43)} $$ \noindent Therefore, when the expectation value
of an exact operator $M$ in a state $\vert \gamma \rangle$ does
not vanish, namely,

$$
\begin{array}{lll}
\langle \gamma \vert M \vert \gamma \rangle \neq 0,
\end{array}
\eqno{(3.44)} $$ \noindent it is an indication that the state
$\vert \gamma \rangle$ is unphysical,

$$
\begin{array}{lll}
Q_B \vert \gamma \rangle \neq 0.
\end{array}
\eqno{(3.45)} $$ \noindent We are going to exploit this fact in
order to derive (3.29). For this purpose we shall remove $\partial
_{\mu}$ in Eq. (3.41). In momentum space $\partial _{\mu}$ denotes
the momentum transfer so that we differentiate the Fourier
transform of Eq. (3.41) with respect to the momentum transfer and
take the limit of zero momentum transfer. We shall illustrate this
procedure in QED  by starting from the following Ward-Takahashi
identity:

$$
\begin{array}{lll}
(p-q)_{\mu} \Gamma _{\mu} (p,q) = -i (S_F^{-1} (p) - S_F^{-1}
(q)).
\end{array}
\eqno{(3.46)} $$ \noindent We differentiate this identity with
respect to $p_{\mu}$ and then take the limit of $q \rightarrow p$
and obtain

$$
\begin{array}{lll}
\Gamma _{\mu} (p,p) = -i \frac{\partial}{\partial p_{\mu}}
S_F^{-1} (p).
\end{array}
\eqno{(3.47)} $$ \noindent This derivation is valid provided that
$\Gamma_{\mu} (p,q)$ does not have a pole at $(p-q)^2=0$.

Now define the spin zero projection of $\delta \bar{\delta}
A_{\mu}$ and denote it by $M_{\mu}$, then apply the above
procedure to a colored field $\Phi^{\alpha}$, then we have

$$
\begin{array}{lll}
\langle p, \beta \vert M_{\mu}^{a} \vert p, \alpha \rangle \propto
R_{\beta\alpha}^{a} \times (kinematical \ factor),
\end{array}
\eqno{(3.48)} $$ \noindent where $\vert p, \alpha \rangle$ denotes
a state involving a quantum of $\Phi^{\alpha}$ with four-momentum
$p$. In this derivation, however, we have assumed the absence of
the massless pole as in QED, and this assumption is expressed by
[22]

$$
\begin{array}{lll}
\delta \bar{\delta} \chi^{a} = 0.
\end{array}
\eqno{(3.49)} $$

\noindent The derivation of (3.48) in the absence of the asymptotic fields
is discussed in Appendix B.

Then we can refer to the argument based on (3.44) and (3.45) and
we conclude that the quantum of the colored field $\Phi^{\alpha}$
is confined. For a color singlet field we have $R=0$ and its
quantum is not confined just as in the case of (3.36). Thus we may
conclude that Eq. (3.49) is a sufficient condition for c.c. in the
sense of (3.29).

At this stage we quote some examples in which this condition is
$not$ satisfied. The first example is $an\ abelian\ gauge\ theory$
represented by QED. In this case the quanta of the FP ghost fields
are free so that they cannot form bound states with charged
particles, and charge confinement cannot be realized. We can also
give an alternative explanation: In this case we can easily derive

$$
\begin{array}{lll}
i \delta \bar{\delta} \chi = - \beta,
\end{array}
\eqno{(3.50)} $$ \noindent and the condition (3.49) is not
satisfied.

Another example is found $when\ a\ certain\ gauge\ symmetry\ is \
spontaneously\ broken$. In analogy with Eq. (3.46) we may express
the identity (3.41) for the quark field in momentum space as

$$
\begin{array}{lll}
(p-q)_{\mu} V_{\mu}^{a} (p,q) = -ig T_{\alpha\beta} ^{a} \left [
S_F^{-1} (p, \alpha) - S_F^{-1} (q, \beta)\right].
\end{array}
\eqno{(3.51)} $$ \noindent Assume that the gauge symmetry
corresponding to the group index $a$ is broken and that the quark
masses for the colors $\alpha$ and $\beta$ are non-degenerate. In
the limit $q\rightarrow p$ the r.h.s. does not vanish because of
the mass difference,

$$
\begin{array}{lll}
S_F^{-1} (p, \alpha) - S_F^{-1} (p, \beta) \neq 0.
\end{array}
\eqno{(3.52)} $$ \noindent This implies that $V_{\mu}^{a} (p,q)$
should develop a massless pole. Physically the appearance of this
pole is a signal that the Nambu-Goldstone boson has emerged in the
form

$$
\begin{array}{lll}
\delta \bar{\delta} \chi ^{a} \neq 0, \Box \delta \bar{\delta}
\chi ^{a} = 0,
\end{array}
\eqno{(3.53)} $$ \noindent which violates the condition (3.49).

These two cases serve to explain why the electroweak interactions
do not confine any particles. Indeed, the electroweak gauge theory
is constructed on the gauge group $SU(2) \times U(1)$, but this
symmetry is spontaneously broken and reduced to U(1) corresponding
to the electromagnetic gauge symmetry. Thus we are left with only
an abelian gauge symmetry and this explains why the electroweak
interactions do not lead us to confinement.

\subsection{Superconvergence relation}

Quite independently of (3.49) a condition for gluon confinement
has been derived from the study of the renormalization constant of
the color gauge field.

Let us introduce the propagator of the color gauge field,

$$
\begin{array}{lll}
\langle A_{\mu} ^{a} (x) A_{\nu} ^{b} (y) \rangle = \frac{-i}{(2 \pi)^4}\delta_{ab}
\int d^4 k e^{i k \cdot (x-y)} D_F (k)_{\mu\nu},
\end{array}
\eqno{(3.54)} $$ \noindent where

$$
\begin{array}{lll}
D_F (k)_{\mu\nu} = (\delta_{\mu\nu} - \frac{k_{\mu}
k_{\nu}}{k^2-i\epsilon}) D(k^2) + \alpha \frac {k_{\mu}
k_{\nu}}{(k^2-i\epsilon)^2}.
\end{array}
\eqno{(3.55)} $$ \noindent We introduce the Lehmann representation
[23] for $D(k^2)$ by

$$
\begin{array}{lll}
D(k^2) = \int dm^2 \frac{\rho (m^2)}{k^2+m^2-i\epsilon},
\end{array}
\eqno{(3.56)} $$ \noindent then Lehmann's theorem gives $$
\begin{array}{lll}
Z_3^{-1} = \int dm^2 \rho (m^2),
\end{array}
\eqno{(3.57)} $$ \noindent where $Z_3$ is the renormalization
constant of the color gauge field. Oehme and Zimmermann [24] have
studied the structure of this propagator in the Landau gauge and
have proved the superconvergence relation

$$
\begin{array}{lll}
Z_3^{-1} = \int dm^2 \rho (m^2) = 0,
\end{array}
\eqno{(3.58)} $$ \noindent when the number of quark flavors $N_f$
is less than 10, namely, in the case 1) of Subsection 5.3 as we
shall see later. This proof is based on the renormalization group
(RG) method, and we shall come back to this method later. Then it
was recognized by one of the present authors (KN) and also by
Oehme that gluon confinement follows from this superconvergence
relation [25,26].

Thus it turns out to be an important problem to clarify the
relationship between these two conditions (3.49) and (3.58) and we
shall study it in the next section.

\section{Renormalization Group and Color Confinement}

In the preceding section we have obtained two kinds of conditions
for c.c., and in this section we elucidate their relationship in
order to find the most fundamental condition.

\subsection{Renormalization group}

The infinitesimal operator of the RG is given by the differential
operator

$$
\begin{array}{lll}
{\cal D} = \mu \frac{\partial}{\partial \mu} + \beta (g)
\frac{\partial}{\partial g} - 2 \alpha \gamma_{V} (g,\alpha)
\frac{\partial}{\partial \alpha},
\end{array}
\eqno{(4.1)} $$ \noindent where $\mu$ denotes the renormalization
point, $\alpha$ the gauge parameter, and $\gamma _{V}$ the
anomalous dimension of the color gauge field. An element of the RG
may be expressed as

$$
\begin{array}{lll}
R(\rho) = \exp (\rho {\cal D}),
\end{array}
\eqno{(4.2)} $$ \noindent where $\rho$ is the parameter of the RG
and we have the composition law of the group, $$
\begin{array}{lll}
R(\rho) \cdot R(\rho^{'})  = R(\rho + \rho^{'}).
\end{array}
\eqno{(4.3)} $$ \noindent Let $Q$ be a function of $g$, $\alpha$
and $\mu$, and we define the running $Q$ by

$$
\begin{array}{lll}
\overline{Q} (\rho) &=& \exp (\rho {\cal D}) \cdot
Q(g,\alpha,\mu)\\ &=&Q(\bar{g}(\rho), \bar{\alpha} (\rho),
\bar{\mu}(\rho)),
\end{array}
\eqno{(4.4)} $$ \noindent with the initial condition

$$
\begin{array}{lll}
\overline{Q} (0) = Q.
\end{array}
\eqno{(4.5)} $$ \noindent Then introduce Green's function
$G(p_{i};g,\alpha,\mu)$ and let its anomalous dimension be $\gamma
(g,\alpha)$, then we have a RG equation:

$$
\begin{array}{lll}
[{\cal D} + \gamma (g, \alpha)] G(p_i;g,\alpha,\mu) = 0.
\end{array}
\eqno{(4.6)} $$ \noindent Its running version defined by $$
\begin{array}{lll}
\overline{G} (\rho) = \exp (\rho {\cal D}) \cdot
G(p_i;g,\alpha,\mu).
\end{array}
\eqno{(4.7)} $$ \noindent satisfies the following equation: $$
\begin{array}{lll}
\frac{\partial}{\partial \rho} \overline{G} (\rho) = -
\bar{\gamma} (\rho) \overline{G} (\rho).
\end{array}
\eqno{(4.8)} $$ \noindent Its integral is given by $$
\begin{array}{lll}
G(p_i;g,\alpha,\mu) = \exp [\int_{0}^{\rho} d\rho^{'} \bar{\gamma}
(\rho^{'})]\cdot G(p_i;\bar{g} (\rho),\bar{\alpha} (\rho),
\bar{\mu} (\rho)).
\end{array}
\eqno{(4.9)} $$

The RG provides us with the relationship between renormalized and
unrenormalized expressions. In order to give a finite value to an
unrenormalized expression we have to introduce a cut-off
$\Lambda$, then higher order corrections tend to decrease for
momentum-transfer beyond the cut-off $\Lambda$. Then we may assume
that the running coupling constant $\bar{g} (\rho)$ tends to the
unrenormalized one or the bare one $g_0$ in the limit $\rho
\rightarrow \infty$,

\begin{eqnarray*}
\hspace{3cm} \lim_{\rho \rightarrow \infty} \bar{g} (\rho) = g_0
\end{eqnarray*}
\noindent provided that the cut-off $\Lambda$ is kept finite. In
the RG approach we usually formulate the initial conditions by
keeping $\Lambda$ finite, then we can take the limit $\rho
\rightarrow \infty$ in Eq. (4.9) to obtain

$$
\begin{array}{lll}
G(p_i;g,\alpha,\mu) = \exp [\int_0^{\infty} d\rho^{'} \bar{\gamma}
(\rho^{'})] \cdot G^{(0)} (p_i;g_0,\alpha_0,\infty),
\end{array}
\eqno{(4.10)} $$ \noindent where $G^{(0)}$ denotes the
unrenormalized Green function and the factor

$$
\begin{array}{lll}
Z = \exp [-\int_0^{\infty} d\rho^{'} \bar{\gamma} (\rho^{'})].
\end{array}
\eqno{(4.11)} $$ \noindent gives the renormalization constant of
Green's function $G$. In particular the gluon propagator $D(k^2)$
satisfies

$$
\begin{array}{lll}
R(k^2;g,\alpha,\mu) = k^2 D(k^2;g,\alpha,\mu) = 1, \ for\
k^2=\mu^2,
\end{array}
\eqno{(4.12)} $$ \noindent since this is precisely the definition
of the renormalization point $\mu$. Replacing $G$ by $R$ in (4.9)
we find

$$
\begin{array}{lll}
R(k^2;g,\alpha,\mu) = \exp [2\int_0^{\rho} d\rho^{'}
\bar{\gamma}_{V} (\rho^{'})] \cdot R(k^2;\bar{g} (\rho),
\bar{\alpha} (\rho), \bar{\mu} (\rho)).
\end{array}
\eqno{(4.13)} $$ \noindent Then, by putting $k^2=\bar{\mu} ^2
(\rho)$ in (4.13) and referring to (4.12) we find $$
\begin{array}{lll}
R(\bar{\mu}^2 (\rho);g,\alpha,\mu) = \exp [2\int_0^{\rho}d\rho^{'}
\bar{\gamma}_{V} (\rho^{'})].
\end{array}
\eqno{(4.14)} $$ \noindent Now insert the Lehmann representation
(3.56) into the l.h.s. of Eq. (4.14) and take the limit $\rho
\rightarrow \infty$ or $\bar{\mu} ^2 (\rho) = \mu^2 \exp (2\rho)
\rightarrow \infty$, and we find, for a finite cut-off, the
relation

$$
\begin{array}{lll}
\int dm^2 \rho (m^2) = \exp [2\int_0^{\infty}d\rho^{'}
\bar{\gamma}_{V} (\rho^{'})].
\end{array}
\eqno{(4.15)} $$ \noindent Combing this relation with Lehmann's
theorem (3.57), we find

$$
\begin{array}{lll}
Z_3^{-1} = \exp [2\int_0^{\infty}d\rho^{'} \bar{\gamma}_{V}
(\rho^{'})],
\end{array}
\eqno{(4.16)} $$ \noindent and it satisfies the RG equation $$
\begin{array}{lll}
({\cal D} + 2 \gamma _{V}) Z_3^{-1} = 0.
\end{array}
\eqno{(4.17)} $$ \noindent In the cut-off theory we first take the
limit $\rho \rightarrow \infty$ and then $\Lambda \rightarrow
\infty$, but in what follows we invert the order of limiting
procedures by taking the limit $\Lambda \rightarrow \infty$ first.
Thus some of the initial conditions introduced in the cut-off
theory are not necessarily satisfied as we shall see in what
follows.

\subsection{Conditions for color confinement and their
relationship}

  We shall go back to the condition (3.49). Since
  $i\delta \bar{\delta} \chi^{a}$ is a free massless field we may
assume that its most general form is given by

$$
\begin{array}{lll}
i\delta \bar{\delta} \chi^{a} =- C \beta ^{a}.
\end{array}
\eqno{(4.18)} $$ \noindent In QED we have $C=1$ and confinement
requires $C=0$. Now we have to study how to determine the
coefficient $C$, and we start from the following relation based on
Eq. (3.34):

$$
\begin{array}{lll}
\langle i \delta \bar{\delta} \chi^{a} (x), \chi ^{b} (y) \rangle
= C \delta _{ab} D_F (x-y).
\end{array}
\eqno{(4.19)} $$ \noindent The field $\chi$ is a complicated
asymptotic field, however, and we shall express this relation in
terms of Heisenberg operators. For this purpose we shall consider
the two-point function

$$
\begin{array}{lll}
\langle i \delta \bar{\delta} A_{\mu} ^{a} (x), A_{\nu} ^{b} (y)
\rangle,
\end{array}
\eqno{(4.20)} $$ \noindent then because of Eq. (3.15) the most
general form of its Fourier transform can be expressed as

$$
\begin{array}{lll}
(\delta_{\mu\nu} - \frac{k_{\mu} k_{\nu}}{k^2-i\epsilon}) \int
dm^2 \frac{\sigma (m^2)}{k^2+m^2-i\epsilon} + C
\frac{k_{\mu}k_{\nu}}{k^2-i\epsilon}.
\end{array}
\eqno{(4.21)} $$ \noindent The second term corresponds to the
contributions of the spin zero massless particles and we
immediately obtain

$$
\begin{array}{lll}
\partial _{\mu} \langle i \delta \bar{\delta} A_{\mu}^{a} (x),
A_{\nu}^{b} (y) \rangle = i \delta_{ab} C \partial _{\nu} \delta
^4 (x-y).
\end{array}
\eqno{(4.22)} $$ \noindent Furthermore, because of (3.15) the
l.h.s. can be cast in the form of an equal-time commutator,

$$
\begin{array}{lll}
\delta (x_0-y_0) \langle 0 \vert \left[i\delta \bar{\delta}
A_0^{a} (x), A_j^{b} (y)\right] \vert 0 \rangle& = &i \delta_{ab}
C \partial _j \delta ^4 (x-y),\\&& (j=1,2,3).
\end{array}
\eqno{(4.23)} $$ \noindent As has been shown in [27,28] this
constant $C$ satisfies an RG equation

$$
\begin{array}{lll}
({\cal D} - 2 \gamma _{FP}) C = 0.
\end{array}
\eqno{(4.24)} $$ \noindent In order to study the equal-time
commutator (4.23) we shall go back to the unrenormalized version,
then
 $$
\begin{array}{lll}
i\delta \bar{\delta} A_{\nu} ^{(0)} = \partial _{\mu} A_{\mu\nu}
^{(0)} + g_0 \partial _{\mu} (A_{\mu} ^{(0)} \times A_{\nu}
^{(0)}) + g_0 J_{\nu} ^{(0)},
\end{array}
\eqno{(4.25)} $$ \noindent where $A_{\mu\nu} = \partial_{\mu}
A_{\nu} - \partial _{\nu} A_{\mu}$ and the superscript (0) is
attached to unrenormalized expressions. When we insert this
expression into (4.23) we find that the only surviving term is
given by

$$
\begin{array}{lll}
&&\delta (x_0 - y_0) \langle 0 \vert \left[i \delta \bar{\delta}
A_0^{a} (x)^{(0)}, A_j^{b} (y) ^{(0)}\right] \vert 0 \rangle\\
&=&\delta (x_0 - y_0) \langle 0 \vert \left[\partial _k A_{k0}^{a}
(x)^{(0)}, A_j^{b} (y) ^{(0)}\right] \vert 0 \rangle\\ &=& i
\delta_{ab} a^{(0)} \partial_j \delta^4 (x-y),
\end{array}
\eqno{(4.26)} $$ \noindent where $a^{(0)}$ is a parameter which
depends on the normalization of the gauge field. In the
unrenormalized version $a^{(0)}=1$ and in the renormalized version
$a= Z_3^{-1}$. So we find in the unrenormalized version

$$
\begin{array}{lll}
C^{(0)}=a^{(0)}.
\end{array}
\eqno{(4.27)} $$ \noindent Therefore, in the cut-off theory the
boundary condition for $C$ is given by

\begin{eqnarray*}
\hspace{3cm} \lim _{\rho \rightarrow \infty} (\overline{C} (\rho) -
\bar{a}
(\rho)) = 0. \hspace{7.3cm} (4.28)
\end{eqnarray*}
\noindent The RG equations for $\overline{C}$ and $\bar{a}$ are
given, respectively, by

$$
\begin{array}{lll}
\frac{\partial}{\partial \rho} \overline{C} (\rho) = 2
\bar{\gamma} _{FP} (\rho) \overline{C} (\rho),
\end{array}
\eqno{(4.29)} $$

$$
\begin{array}{lll}
\frac{\partial}{\partial \rho} \overline{a} (\rho) = -2
\bar{\gamma} _{V} (\rho) \overline{a} (\rho).
\end{array}
\eqno{(4.30)} $$ \noindent $\gamma_{FP}$ denotes the anomalous
dimension of the FP ghost fields. The formal solution of the RG
equation for $\overline{C} (\rho)$ satisfying the boundary
condition (4.28) is given in terms of $\overline{a} (\rho)$ by

$$
\begin{array}{lll}
\overline{C} (\rho) &=& \bar{a} (\rho) - 2 \int_{\rho}^{\infty}
d\rho ^{'} (\bar{\gamma} _{V} (\rho^{'}) + \bar{\gamma} _{FP}
(\rho^{'})) \bar{a} (\rho^{'})\\ && \times \exp \left[ -2
\int_{\rho}^{\rho^{'}} d \rho^{''} \bar{\gamma}_{FP}
(\rho^{''})\right].
\end{array}
\eqno{(4.31)} $$ \noindent After obtaining this formal solution we
then let the cut-off $\Lambda$ tend to $\infty$. As a consequence
the boundary condition (4.28) is not necessarily satisfied since
we are changing the order of limiting procedures.

Assume that the superconvergence relation (3.58) is satisfied so
that $\bar{a} (\rho)$ vanishes identically, then we conclude from
Eq. (4.31) that $C=\overline{C} (0)$ vanishes also. In other
words, the condition (3.49) or $C=0$ follows from (3.58).
Therefore, the superconvergence relation $Z_3^{-1}=0$ is the most
fundamental condition for c.c., and it will be referred to as the
CCC. This condition was first recognized as the condition for
gluon confinement in the Landau gauge [27] since $Z_3$ had been
known only in this gauge, but later this result was extended to
other gauges and also to other colored particles [28].

It should be mentioned that although we have started from (4.28)
the difference $\overline{C} (\rho) - \bar{a} (\rho)$ does not
necessarily vanish since $\overline{C} (\rho)$ and $\bar{a}
(\rho)$ satisfy different RG equations. In particular, $C-Z_3^{-1}
= \overline{C} (0) - \bar{a} (0)$ represents the contributions of
the so-called Goto-Imamura-Schwinger term [29,30]. When
$Z_3^{-1}=0$, however, this term should also vanish as we have
remarked above [31]. We also have

$$
\begin{array}{lll}
&&\delta (x_0-y_0) \langle 0 \vert \left[ \dot{A}_{\mu} ^{a} (x),
A_{\nu} ^{b} (y) \right] \vert 0 \rangle \\&=& -i \delta _{ab}
\delta _{\mu\nu} Z_3^{-1} \delta^4(x-y) = 0,
\end{array}
\eqno{(4.32)} $$ \noindent and hence the vanishing of the
Goto-Imamura-Schwinger term seems plausible as has been proved
otherwise.

We conclude that c.c. is realized when $Z_3^{-1} = 0$, although
originally only gluons were considered to be confined. This CCC
implies $C=0$, but we already know that $C\neq 0$ for abelian and
broken non-abelian gauge symmetries. Therefore, in these cases
$Z_3^{-1}$ cannot vanish either.

\section{\bf Realization of Color Confinement}

In the preceding section we have obtained a simple but generic
condition for c.c., then a natural question is raised of how to
evaluate the renormalization constant $Z_3$ so that we know
precisely in which theories c.c. is realized. It so happens that
$Z_3$ can be evaluated exactly in QCD and we can readily check the
CCC [28,32], but before entering this subject we have to discuss a
more fundamental subject.


\subsection{Gauge-independence of the concept of color
confinement}

In QED the renormalization constant $Z_3$ does not depend on the
gauge parameter and is hence gauge-independent. It is not the case
in QCD, however, and it is a function of the gauge parameter
$\alpha$ and the gauge coupling constant $g$. Then, what is the
significance of the CCC since $Z_3^{-1}$ might vanish in certain
gauges but not in others?

In order to examine this question we first stress that the concept
of color confinement is gauge-independent. When the condition
(3.17) is satisfied, the only observable particles are hadrons,
namely, color singlet bound states of quarks and gluons and they
are represented by color singlet, and hence BRS invariant,
composite operators when the LSZ reduction formula is applied.
Then the only observable S matrix elements are the transition
amplitudes among hadronic states.

Bearing this in mind we introduce the concept of the equivalence
class of gauges [28,32]. Let us consider a class of Lagrangian
densities $\{{\cal L}_{\alpha}\}$ representing a gauge theory such
as QCD. Assume that all the members of this set are BRS invariant,
$$
\begin{array}{lll}
\delta {\cal L}_{\alpha}=0 ,
\end{array}
\eqno{(5.1)} $$ \noindent and further that the difference between
any two members are exact, namely,
 $$
\begin{array}{lll}
\Delta {\cal L}= {\cal L}_{II}-{\cal L}_{I}= \delta {\cal M} ,
\end{array}
\eqno{(5.2)} $$

\noindent then this set $\{ {\cal L}_{\alpha}\}$ is called an
equivalence class of gauges. Lagrangian densities corresponding to
different choices of $\alpha$ in (3.1) belong to the same
equivalence class.

We introduce Green's functions in two gauges of the same class,
then they are related to one another through the Gell-Mann-Low
relation [33]: $$
\begin{array}{lll}
\langle A(x_1)B(x_2)\cdots \rangle_{II} = \langle
A(x_1)B(x_2)\cdots exp(i\Delta S)\rangle_{I}  ,
\end{array}
\eqno{(5.3)} $$

\noindent where $A(x_1)$, $B(x_2)$, $\cdots$ are local operators,
and

$$
\begin{array}{lll}
\Delta S = \int d^4x \Delta {\cal L} = \int d^4x \delta {\cal M} .
\end{array}
\eqno{(5.4)} $$ \noindent In Eq. (5.3) we assume the convergence
of the series expansion in powers of $\Delta S$.

In particular, when all the local operators are BRS invariant,

$$
\begin{array}{lll}
\delta A=\delta B=\cdots =0 ,
\end{array}
\eqno{(5.5)} $$

\noindent we have, as a result of Eq. (3.43), the equality

$$
\begin{array}{lll}
\langle A(x_1)B(x_2)\cdots \rangle_{II} = \langle
A(x_1)B(x_2)\cdots \rangle_{I}
\end{array}
\eqno{(5.6)} $$

\noindent subject to the convergence of the power series mentioned
above. Thus Green's functions involving only BRS invariant
operators do not depend on the choice of the gauge within the same
equivalence class. When these composite operators represent
hadrons we can apply the LSZ reduction formula [20] to such
Green's functions to obtain the gauge-independent transition
amplitudes for hadronic reactions [34,35,36].

The unitarity condition for the BRS invariant S matrix between two
hadronic states $\vert a\rangle$ and $\vert b\rangle$ is given by

\begin{eqnarray*}
\hspace{2.0cm} \langle b\vert   a\rangle =\langle b   \vert
S^{\dagger}S   \vert a  \rangle =\sum_n  \langle b   \vert
S^{\dagger}\vert n\rangle \langle n\vert S \vert a  \rangle
,\hspace{5.5cm} (5.7)
\end{eqnarray*}

\noindent and a similar one for $SS^{\dagger}$. Color confinement
is realized when the sum over intermediate states is saturated by
hadronic states alone. Since the hadronic S matrix elements are
gauge-independent, c.c. in one gauge automatically prevails in
other gauges of the same equivalence class. This saturation of
intermediate states by hadronic states could be employed as an
alternative interpretation of c.c. since this statement is made in
terms of observable hadrons alone. We shall come back to this new
interpretation later in connection with the Meissner-like effect
in the present approach.

\subsection{Evaluation of $Z_3^{-1}$}

Next we shall proceed to evaluation of $Z_3^{-1}$ and for this
purpose we shall study the RG equations for the running parameters
$\bar{g} (\rho)$, $\bar{\alpha} (\rho)$ and $\bar{\mu} (\rho)$,

$$
\begin{array}{lll}
\frac{d}{d \rho} \bar{g} (\rho) = \bar{\beta} (\rho),
\end{array}
\eqno{(5.8a)} $$

$$
\begin{array}{lll}
\frac{d}{d \rho} \bar{\alpha} (\rho) = -2 \bar{\alpha} (\rho)
\bar{\gamma} _{V} (\rho),
\end{array}
\eqno{(5.8b)} $$

$$
\begin{array}{lll}
\frac{d}{d \rho} \bar{\mu} (\rho) = \bar{\mu} (\rho).
\end{array}
\eqno{(5.8c)} $$ \noindent First, we shall define their asymptotic
values by

$$
\begin{array}{lll}
\bar{g} (\infty) = g_{\infty}, \bar{\alpha} (\infty) =
\alpha_{\infty},\bar{\mu} (\infty) = \infty.
\end{array}
\eqno{(5.9)} $$ \noindent Asymptotic freedom (AF) is characterized
by

$$
\begin{array}{lll}
g_{\infty} = 0,
\end{array}
\eqno{(5.10)} $$ \noindent and it is realized for $N_f \le 16$.

By integrating (5.8b) we find

$$
\begin{array}{lll}
\ln \frac{\alpha_{\infty}}{\alpha} = -2 \int_0^{\infty} d\rho
\bar{\gamma}_{V} (\rho),
\end{array}
\eqno{(5.11)} $$ \noindent or

 $$
\begin{array}{lll}
Z_3^{-1} = \exp [2 \int_0^{\infty} d\rho \bar{\gamma}_{V} (\rho)]=
\frac{\alpha}{\alpha_{\infty}}.
\end{array}
\eqno{(5.12)} $$ \noindent Thus evaluation of $Z_3^{-1}$ reduces
to that of $\alpha_{\infty}$. If we identify $\alpha_{\infty}$
with the unrenormalized gauge parameter this is a
rather trivial relation, but it is not trivial that
$\alpha_{\infty}$ assumes only three possible values $$
\begin{array}{lll}
\alpha_{\infty}= -\infty,  0  ,  \alpha_0 ,
\end{array}
\eqno{(5.13)} $$

 \noindent where $\alpha_0$ is a numerical
constant which depends only on the number of quark flavors as we
shall see in the next subsection.

\subsection{Evaluation of $\alpha_{\infty}$}

Evaluation of $\alpha_{\infty}$ has been published elsewhere
[17,28,32] so that we shall be brief in what follows.

First, we introduce the $\beta$ function and the anomalous
dimension $\gamma_{V}$ as series expansions in powers of the
coupling constant:

$$
\begin{array}{lll}
\beta (g) = g^3 (\beta_0 + \beta_1 g^2 + \cdots),
\end{array}
\eqno{(5.14)} $$

$$
\begin{array}{lll}
\gamma_{V} (g,\alpha) = g^2 (\gamma_0 (\alpha) + \gamma_1 (\alpha)
g^2 + \cdots),
\end{array}
\eqno{(5.15)} $$ \noindent where

$$
\begin{array}{lll}
\gamma_{0} (\alpha) &=& \gamma_{00} + \gamma_{01} \alpha ,\\
\gamma_{1} (\alpha) &=& \gamma_{10} + \gamma_{11} \alpha + \gamma
_{12} \alpha ^2,\\ && \vdots
\end{array}
\eqno{(5.16)} $$ \noindent The lowest order coefficients are given
by

$$
\begin{array}{lll}
\beta_{0} &=& - \frac{1}{32 \pi^2} (22- \frac{4}{3} N_f),\\
\gamma_{00} &=& - \frac{1}{32 \pi^2} (13- \frac{4}{3} N_f), \
\gamma_{01} = \frac{3}{32 \pi^2} > 0.\\
\end{array}
\eqno{(5.17)} $$ \noindent When $\beta_0$ is negative, namely,
when $N_f \le 16$, AF is realized, and we assume it in what
follows. Then for large values of $\rho$ we obtain

$$
\begin{array}{lll}
\bar{g} ^2 (\rho) \approx \frac{1}{b\rho}. \ (b=-2 \beta_0 >0)
\end{array}
\eqno{(5.18)} $$ \noindent In order to check the convergence of
the integral (5.11) we study the behavior of its integrand for
large values of $\rho$ by expanding it in powers of $\bar{g}^2$,
and we can easily verify the convergence when and only when

$$
\begin{array}{lll}
\alpha_{\infty} = \alpha_0,
\end{array}
\eqno{(5.19)} $$ \noindent where $\alpha_0$ is defined by

$$
\begin{array}{lll}
\gamma_0 (\alpha_0) = \gamma_{00} + \gamma_{01} \alpha_0 = 0.
\end{array}
\eqno{(5.20)} $$ \noindent When the integral is divergent we
obtain

$$
\begin{array}{lll}
\alpha_{\infty} = - \infty, \ 0.
\end{array}
\eqno{(5.21)} $$ \noindent Apparently, the CCC (3.58) is satisfied
when $\alpha_{\infty} = -\infty$ because of the sum rule (5.12),
and we have to check if this case is actually realized in QCD. In
order to check which of the three possible values of
$\alpha_{\infty}$ in (5.13) is realized we have to study Eq.
(5.8b) closely with the help of AF. In what follows we shall quote
only the results.

\noindent Case 1)   $\gamma_{00} < 0$, $\alpha_0 > 0$ ($N_f < 10$)

$$
\begin{array}{lll}
\alpha_{\infty} = \left \{ \begin{array}{c}  \alpha_0, \ \ for \
\alpha > 0,\\
 0, \ \ for \ \alpha = 0, \\
-\infty,\ \ for \ \alpha < 0.
\end{array} 
\right .
\end{array}
\eqno{(5.22)} $$ \noindent Case 2)   $\gamma_{00} > 0$, $\alpha_0
< 0$ ($10\leq N_f \leq 16$)

$$
\begin{array}{lll}
\alpha_{\infty} = \left \{ \begin{array}{c}  0, \ \
for \ \alpha >
\alpha_0 + h(g^2),\\
 \alpha_0, \ \ for \ \alpha = \alpha_0 + h(g^2),\\ -\infty,
\ \ for \ \alpha < \alpha_0 + h(g^2),
\end{array}
\right .
\end{array}
\eqno{(5.23)} $$ \noindent where

$$
\begin{array}{lll}
\alpha =  \alpha_0 + h(g^2) = \alpha_0 + g^2 (h_0+ h_1 g^2 +
\cdots )
\end{array}
\eqno{(5.24)} $$ \noindent is a special solution of the following
equation:

$$
\begin{array}{lll}
\frac{d\alpha}{dg} =  -2 \alpha \frac{\gamma_{V} (g,
\alpha)}{\beta (g)}.
\end{array}
\eqno{(5.25)} $$ \noindent Thus in both cases $\alpha_{\infty} =
-\infty$ is realized when the gauge parameter and the gauge
coupling constant are properly chosen and consequently the CCC is
satisfied.

A domain in the $(\alpha, g^2)$ half-plane corresponding to
$\alpha_{\infty}$ is denoted by $D(\alpha_{\infty})$, then this
half-plane is covered by $\overline{D} (-\infty)$, $\overline{D}
(0)$, $\overline{D} (\alpha_0)$, where the bar denotes closure.
The above arguments show that c.c. is realized in $D(-\infty)$,
but what happens in the other two domains? Formally we can apply
the argument on the gauge-independence of the concept of c.c.,
namely, for two different values of $\alpha$ we have

$$
\begin{array}{lll}
\Delta {\cal L} = \frac{1}{2} (\Delta \alpha) B \cdot B =
-\frac{i}{2} (\Delta \alpha) \delta (\bar{c} \cdot B),
\end{array}
\eqno{(5.26)} $$ \noindent or

$$
\begin{array}{lll}
{\cal M} = -\frac{i}{2} (\Delta \alpha) (\bar{c} \cdot B),
\end{array}
\eqno{(5.27)} $$ \noindent so that we can utilize the formula
(5.6) to show that c.c. prevails in the other two domains. In this
case $Z_3^{-1}=0$ is a sufficient condition but not a necessary
one for c.c. since it is not satisfied in the two domains $D(0)$
and $D(\alpha_0)$. The identity (5.3) is valid only when the
series expansion in powers of $\Delta S$ is convergent. In the
present case we have to check the convergence of the power series
in $\Delta \alpha$. Then Eq. (4.9) shows within the framework of
the RG approach that this implies the convergence of the power
series in $\Delta \bar{\alpha} (\rho)$. Let us consider, for
instance, the case 1), then the line $\alpha=0$ is the border
between $D(-\infty)$ and $D(\alpha_0)$. Let us introduce $\alpha_1
< 0$ and $\alpha_2 > 0$, then even when $\vert \Delta \alpha \vert
= \vert \alpha_2 - \alpha_1 \vert \ll 1$, $\Delta \bar{\alpha}
(\rho)$ tends to $\infty$ in the limit $\rho \rightarrow \infty$
and the convergence of the power series turns out to be doubtful
thereby suggesting that $\alpha=0$ would be a branch point of
Green's functions involving BRS variant operators. This, in turn,
means that such Green's functions would be multi-valued functions
of $\alpha$, Then a question is raised of whether Green's
functions involving only BRS invariant operators be also
multi-valued. When this were the case c.c. might be realized only
in the domain $D(-\infty)$, so that $Z_3^{-1}=0$ would represent a
necessary and sufficient condition for c.c., but it has not been
clarified yet if it would really be the case.

We may close this section by concluding that c.c. is realized in
QCD provided that color symmetry is not spontaneously broken and
AF is valid.

\section{\bf Consequences of the CCC}

In this section we shall show how the CCC, $Z_3^{-1}=0$, is
related to other interpretations of confinement based mainly on the
dual Meissner effect. Since the starting point of other
interpretations are quite distinct from that of the present
approach it is not easy to compare the basic formulations for the
purpose of clarifying their relationships. Therefore, we shall try
to compare the consequences of our approach with those of the
others.

An attempt has been made, however, by one of the present authors
(M.C.) and Kobayashi [37] to clarify the relationship between the
Seiberg formulation [38,39,40] and the present superconvergence
rule in the Landau gauge by studying the criteria for confinement
expressed in terms of the $\beta$ function and the anomalous
dimension of the gauge field.

\subsection{An intuitive interpretation of the CCC}

Let us consider a dielectric medium and put a positive test charge
inside, then negative charges are attracted and positive ones are
repelled by it. Therefore, it induces a new charge distribution in
the medium and the total charge inside a sphere of radius $r$
around the test charge is a function of $r$. It is denoted by
$\overline{e}(r)$ and called the running charge.

In classical physics the vacuum means the empty mathematical space
or the void, but in quantum physics the physical vacuum is a kind
of medium with a rich structure and has to be distinguished from
the classical vacuum or the mathematical space. The dielectric
constant is defined relative to one of the vacua. Now let us
regard the physical vacuum as a dielectric medium and call the
test charge $e_0=\bar{e} (0)$ the bare charge and the total charge
inside a sphere of a sufficiently large radius $e=\bar{e}
(\infty)$ the renormalized charge.

We then introduce the dielectric constant of the physical vacuum
$\epsilon$ relative to the mathematical one and write down the
static Coulomb potential between two electrons in two alternative
ways,

$$
\begin{array}{lll}
V(r) = \frac{e^2}{4\pi r} = \frac{e_0^2}{4 \pi \epsilon r}.
\end{array}
\eqno{(6.1)} $$ \noindent We also introduce the renormalization
constant $Z_3$ of the electromagnetic field, then we find

$$
\begin{array}{lll}
e_0^2 = Z_3^{-1} e^2,
\end{array}
\eqno{(6.2)} $$ \noindent Thus we are led to

$$
\begin{array}{lll}
\epsilon =Z_3^{-1} .
\end{array}
\eqno{(6.3)} $$

\noindent At the end of Section 2 we have discussed the two-phase
structure or the emergence of the Meissner effect in the limit
$\epsilon\rightarrow 0$, and now through Eq. (6.3) we find that
this condition is equivalent to the CCC. Furthermore, we find that
this condition along with (6.2) implies $e_0=0$ or AF. In the
extreme case of $\epsilon =0$, a small test charge would attract
an unlimited amount of like charges around it leading the system
into a catastrophic state of infinite charge. Nature would take
safety measures to prevent such a state from emerging, and a
possible resolution to avoid it would be to bring another test
particle of opposite charge. The total charge of the whole system
is then equal to zero and charge confinement would be realized.

 The dielectric constant of the vacuum $\epsilon$ is larger than
 unity, however,
as a consequence of the screening effect due to vacuum
polarization. Quantum mechanically $Z_3^{-1}$ is larger than unity
because of the contributions of charged particles of
positive-definite metric so that the condition $Z_3^{-1}=0$ could
be realized only in the presence of charged particles of
indefinite metric. They do not appear in QED unless they are put
by hand just as the magnetic monopoles in Section 2, however, so
that electric charges cannot be confined in QED. Indeed, this is a
conclusion repeatedly drawn, but QED is a good laboratory for
Gadankenexperiment, however fictitious, to illustrate the
mechanism of confinement.

\subsection{The CCC and the linear potential}

In this subsection we shall give an intuitive argument on a
possible connection between the CCC and Wilson's area law in the
lattice gauge theory [6].

Wilson has formulated the criterion for confinement in terms of a
loop correlation function defined by

$$
\begin{array}{lll}
W[C]=Tr P \exp (ig \int_C A_{\mu} dx_{\mu}),
\end{array}
\eqno{(6.4)} $$ \noindent where $P$ stands for the path ordering.
$W$ creates a tube of flux, along the path $C$, of strength equal
to one quark color charge. a quark-antiquark pair is attached to
both ends of the open path or the path is closed. The average
value of this expression is defined in terms of path integrals in
Euclidean space, and for a large closed path $C$, its average
value is asymptotically proportional to

$$
\begin{array}{lll}
\langle W \rangle \propto \exp (-perimeter),
\end{array}
\eqno{(6.5)} $$ \noindent
or

$$
\begin{array}{lll}
\langle W \rangle \propto \exp (-area).
\end{array}
\eqno{(6.6)} $$ \noindent The perimeter and the area denote those
of the path $C$, respectively. Wilson's criterion for confinement
is the realization of the latter case called the area law.

In evaluating the loop correlation function (6.4) we shall choose
a rectangular contour with a temporal extension $T$ and a spatial
extension $R$, and the effective potential between a heavy quark
and a heavy antiquark is given by

\begin{eqnarray*}
\hspace{3cm}
V[R] = - \lim _{T\rightarrow \infty} \frac{1}{T} \ln W [R\times
T]. \hspace{6.0cm} (6.7)
\end{eqnarray*}

\noindent When the area law (6.6) is obeyed we have the following
linear potential at large distances:

$$
\begin{array}{lll}
V[R] = \sigma R.
\end{array}
\eqno{(6.8)} $$ \noindent The linear potential is no longer valid,
however, when a quark-antiquark pair can be created from the
vacuum, since it is energetically more favorable to split the tube
of flux between the heavy quark pair thereby attaching a light
quark and a light antiquark pair to the two split ends than to
stretch the string indefinitely. We shall realize this point later
in this subsection.

In evaluating the expression (6.7) we shall introduce a cluster
expansion as given by

$$
\begin{array}{lll}
&& \ln \langle Tr P \exp (ig \oint _{C} A_{\mu} dx_{\mu}) \rangle
\\ &=& - \frac{1}{2} g^2 \oint _{C} \oint _{C} \langle A_{\mu} (x)
A_{\nu} (y) \rangle dx_{\mu} dy_{\nu} + \cdots.
\end{array}
\eqno{(6.9)} $$ \noindent Since we are taking the limit
$T\rightarrow \infty$ for the path $C=R\times T$ we may keep only
the temporal path, and we obtain

$$
\begin{array}{lll}
V[R] \propto \int dx_0 D_F (x)_{00},
\end{array}
\eqno{(6.10)} $$ \noindent where $D_F (x)_{\mu\nu}$ is the gluon
propagator in Eq. (3.54), and we may insert the representations
(3.55) and (3.56) into the integrand.

When the CCC is satisfied we may express $D(k^2)$ as

$$
\begin{array}{lll}
D(k^2)= \int dm^2\frac{ \tau (m^2)}{(k^2+ m^2 -i \epsilon)^2} ,
\end{array}
\eqno{(6.11)} $$
\noindent where

$$
\begin{array}{lll}
\tau (m^2) = \int_0^{m^2} dM^2 \rho (M^2) ,
\end{array}
\eqno{(6.12)} $$

\noindent with

$$
\begin{array}{lll}
\tau(0)= \tau(\infty)=0.
\end{array}
\eqno{(6.13)} $$

Now we switch to a rather crude approximation since the concept of
the static potential itself is phenomenological in nature.
Actually $r=0$ is a singularity in the potential. Although the CCC
suppresses the $1/r$ singularity we are still left with a mild
singularity at $r=0$, and this forces us to introduce a crude
approximation. Let us assume that $\vert \tau (m^2)\vert$ has a
maximum at $m^2= \kappa^2$, then because of asymptotic freedom
gauge interactions are stronger at lower energies than at higher
energies and $\kappa$ represents a typical mass of unconfined
systems sharing the same set of quantum numbers with a single
gluon. Thus we may further assume that $\kappa$ is of the order of
or even smaller than the pion mass. Then we may approximate (6.11)
by a smeared dipole type propagator

$$
\begin{array}{lll}
D(k^2) \approx \frac{f}{(k^2+ \kappa^2-i\epsilon)^2} .
\end{array}
\eqno{(6.14)} $$

\noindent Then the static potential $V_{q\overline{q}}(r)$ between
a heavy quark and a heavy antiquark can be evaluated by utilizing
(6.10) as

$$
\begin{array}{lll}
V_{q\overline{q}}(r)&\propto& \frac{1}{\kappa} e^{-\kappa r}\\
&\cong& \frac{1}{\kappa} - r, \  for \ r \ll 1/\kappa.
\end{array}
\eqno{(6.15)} $$

\noindent Thus we have a linear potential in a limited range. The
first constant term may be included in the self energies of the
heavy quarks [41,42]. An unlimited extension of a linear potential
to large distances would lead to too strong a van der Waals force
between hadrons [43] so that the potential should cease to be
linear and fall off at large distances as remarked in the
beginning of this subsection. In this connection it should be
mentioned that the approximate linear potential is a result of but
not the cause of color confinement as we have seen already.

The $q\overline{q}$ interactions discussed here are generated by
exchanging {\it colored objects} and has a relatively long range.
This indicates, however, that vortices connecting a
quark-antiquark pair should have a finite length of the order of
or less than $1/\kappa$. As we shall see in the next subsection the
hadron interactions are generated by exchanging {\it color singlet
hadrons} and are of the short-range type.

\subsection{Hadron interactions and the Meissner-like effect} 

In the Subsection 6.1 we have discussed the emergence of the
two-phase structure and the related dual Meissner effect. This is
expected to take place in the limit of the vanishing dielectric
constant. In real QED, however, it is larger than unity and this
scenario fails to be realized. For instance, let us consider the
interaction between two electrically neutral systems. Then, their
interaction is given by the van der Waals potential $$
\begin{array}{lll}
V_{vdW} (r)\propto r^{-6} .
\end{array}
\eqno{(6.16)} $$
\noindent This shows that the electric field
generated by neutral systems penetrate into the vacuum without any
sharp cut-off indicating the failure of the dual Meissner effect.

In QCD we can elucidate the dynamics of hadrons by making
reference to dispersion relations. It has been clarified by Oehme
[44] that dispersion relations for the scattering of hadrons
remain applicable provided that confinement excludes quarks and
gluons from the physical state vector space. This is precisely one
of the consequences of the CCC, and we can extract necessary
information about hadron interactions from dispersion relations on
the assumption that the complete hadron spectrum can be accounted
for by QCD.

Let us consider the nucleon-nucleon scattering as an example, then the
potential between them is given by the pole contributions in the crossed
channels. The least massive hadron that can be exchanged between them is the
pion, and the resulting interaction is represented by the Yukawa potential,

$$
\begin{array}{lll}
V_Y (r)\propto \frac{e^{-\mu r}}{r} ,
\end{array}
\eqno{(6.17)} $$
\noindent where $\mu$ denotes the pion mass.

When we compare this result with the van der Waals potential we
recognize that the flux of the color gauge field emerging from
color singlet nucleons cannot penetrate into the confining vacuum
leaving {\it no trace of long range forces} and that the
penetration depth is given by the pion Compton wave length. Thus
we notice that the Yukawa mechanism of generating the nuclear
forces bears a strong resemblance to the Meissner effect in the
type II superconductor.

Although we have chosen a rather abstract approach to c.c. on the
basis of BRS symmetry and asymptotic freedom, it shares
essentially the same salient features with other approaches in
that the vacuum allows penetration of the chromoelectric flux from
hadrons only by a finite depth into it.

To conclude we stress that all these characteristic features of
confinement are the consequences of the CCC, namely, $Z_3^{-1}=0$.

\vskip 5mm
{\bf Acknowledgements}
\vskip 3mm
This work is partially supported by the Academy of Finland under
the Project No. 163394. We are grateful to W.-F. Chen and C. Montonen 
for useful comments.


\section*{Appendix A: Derivation of Dighost Bound States from the Condition (3.49)}

In this appendix we shall prove the existence of the dighost bound states on the basis of
the condition (3.49) by utilizing the properties of the representation of the BRS charge [16,17].

Let us introduce a complete set of basis $\{\vert e_i\rangle\}$ in the state vector
space $\cal{V}$ and define the matrix $\eta$ by

$$
\begin{array}{l}
\eta_{ij}= \langle e_i\vert e_j\rangle . 
\end{array}
\eqno{(A.1)} $$

\noindent Then, due to the properties of the inner product,  it is hermitian,

$$
\begin{array}{l}
\eta^{\dagger}=\eta  . 
\end{array}
\eqno{(A.2)} $$

\noindent Provided that $\cal{V}$  is non-degenerate we can always choose the set
of basis so as to satisfy the constraint

$$
\begin{array}{l}
\eta^{2}=1  , 
\end{array}
\eqno{(A.3)} $$

\noindent which defines the standard form of $\eta$.

The representation $t$ of a given linear operator  $T$ with respect to the given
complete set of basis is defined by

$$
\begin{array}{l}
T\vert e_i\rangle = \sum_j \vert e_j\rangle t_{ji}, 
\end{array}
\eqno{(A.4)} $$

\noindent and the hermitian conjugate $T^{\dagger}$ is defined by

$$
\begin{array}{l}
\langle k\vert T\vert l\rangle = \langle l\vert T\vert k\rangle^{*} , 
\end{array}
\eqno{(A.5)} $$

\noindent for an arbitrary pair of states $\vert l\rangle $ and $\vert k\rangle $.
Then the representation of $T^{\dagger}$ denoted by $\tilde{t}$ satisfies the relation

$$
\begin{array}{l}
(\eta t)^{\dagger} = \eta\tilde{t} .
\end{array}
\eqno{(A.6)} $$

The BRS charge $Q_B$ is hermitian and nilpotent as discussed in Section 3, and its representation
$q$ satisfies

$$
\begin{array}{ll}
q^2=0 , & q^{\dagger} = \eta q\eta .
\end{array}
\eqno{(A.7)} $$

\noindent Now we can introduce the Hodge decomposition of $\cal{V}$ as

$$
\begin{array}{l}
{\cal V}={\cal V}_s+{\cal V}_d+{\cal V}_p .
\end{array}
\eqno{(A.8)} $$
 
\noindent The typical members of these subspaces satisfy the following relations:

$$
\begin{array}{rll}
q\vert s\rangle =0,\ q^{\dagger} \vert s\rangle =0 , & \mbox{for} ~\vert s\rangle \in 
{\cal V}_s ~\mbox{(harmonic)} ,\\
&&\\
q\vert d\rangle=0 ,\ q^{\dagger} \vert d\rangle \neq 0 , & \mbox{for} ~\vert d\rangle \in {\cal V}_d ~\mbox{(exact)}, \\
&&\\
q\vert p\rangle\neq 0 ,\ q^{\dagger} \vert p\rangle =0 , & \mbox{for} ~\vert p\rangle \in {\cal V}_p ~\mbox{(co-exact)} .\\
\end{array}
\eqno{(A.9)} $$

\noindent In each subspace we introduce a complete set of basis such as $\{ \vert s_i\rangle\}$, 
$\{ \vert d_i\rangle\}$ and $\{ \vert p_i\rangle\}$ satisfying

$$
\begin{array}{l}
\langle s_i\vert s_j\rangle = \langle d_i\vert p_j\rangle=\delta_{ij} , \\
\langle d_i\vert d_j\rangle = \langle p_i\vert p_j\rangle=0 .
\end{array}
\eqno{(A.10)} $$

\noindent With these preliminaries we shall start from an analogue of (3.34),

$$
\begin{array}{l}
 \langle \chi^a(x), \overline{\beta}^b(y)\rangle= \delta_{ab} D_F(x-y) ,
\end{array}
\eqno{(A.11)} $$ 

\noindent where $\overline{\beta}$ is the asymptotic field of $\overline{B}$. Then, by
making use of (3.43) and (3.11b) we find

$$
\begin{array}{l}
 \langle \overline{\delta}\chi^a(x), c^{b}(y){}^{in}\rangle= i\delta_{ab} D_F(x-y) .
\end{array}
\eqno{(A.12)} $$ 

\noindent When the condition (3.49) is valid we have 

$$
\begin{array}{l}
 \overline{\delta}\chi^a(x)\vert 0\rangle \in {\cal V}_s+{\cal V}_d .
\end{array}
\eqno{(A.13)} $$

\noindent However, $\overline{\delta}\chi^a(x)\vert 0\rangle $ has zero norm as seen
from

$$
\begin{array}{l}
 \langle 0\vert \overline{\delta}\chi^a(x)\overline{\delta}\chi^a(x)\vert 0\rangle 
= \langle 0\vert \overline{\delta}\left(\chi^a(x)\overline{\delta}\chi^a(x)\right)\vert 0\rangle=0 ,
\end{array}
\eqno{(A.14)} $$ 

\noindent where we have used (3.43) for $\overline{Q}_B$.  Hence we conclude

$$
\begin{array}{l}
 \overline{\delta}\chi^a(x)\vert 0\rangle \in {\cal V}_d .
\end{array}
\eqno{(A.15)} $$ 

\noindent The orthogonality conditions (A.10) and (A.12) then imply that $c^{b}(y){}^{in}\vert 0\rangle$
should involve a  non-vanishing co-exact component and hence

$$
\begin{array}{l}
\delta c^{b}(y){}^{in}\vert 0\rangle = -\frac{1}{2}g \left(c(y)\times c(y)\right)^{b,in}
\vert 0\rangle \neq 0 .
\end{array}
\eqno{(A.16)} $$ 

\noindent This establishes the existence of the dighost bound states.

\section*{Appendix B: Proof of the Condition (3.45) for Colored Particles}

When the condition (3.49) is satisfied we can prove (3.44)  and hence
(3.45) for colored particles. For this purpose we have to derive (3.48) by applying
the reduction formula to the identity (3.41). In the absence of the asymptotic fields, however,
we have to devise a new proof. 

We start from the Lehmann representation [23] for a complex scalar field,

$$
\begin{array}{ll}
\Delta_F^{\prime}(x-y) &= \langle 0\vert T\left[\phi (x)\phi^{\dagger} (y)\right]\vert 0\rangle \\
&\\
& =  \int_{m^2}^{\infty}dM^2 \rho(M^2) \Delta_F(x-y,M^2) ,\\
\end{array}
\eqno{(B.1)} $$ 

\noindent where  $m$ denotes the mass of the quantum of this field, and

$$
\begin{array}{l}
\Delta_F(x,M^2) = \frac{-i}{(2\pi)^4}\int d^4p e^{ip\cdot x} \Delta_F(p,M^2) , 
\end{array}
\eqno{(B.2)} $$ 

\noindent where

$$
\begin{array}{l}
\Delta_F(p,M^2) = (p^2+M^2-i\epsilon)^{-1} .
\end{array}
\eqno{(B.3)} $$

There are two alternative possibilities for the nature of the spectral function $\rho (M^2)$:\\

case i)
$$
\begin{array}{l}
\rho (M^2) = \delta (M^2-m^2) +\sigma (M^2) \theta\left[M^2-(m+\mu)^2 \right] ,\ \mu\neq 0 .
\end{array}
\eqno{(B.4)} $$ 

\hspace{1.2cm} In this case there is a pole term in the propagator on the mass shell.\\

case ii)

\hspace{1.2cm} There is no pole term due to infrared singularities and $\mu=0$ in this
case.\\

In the case i) there are asymptotic fields defined by

$$
\begin{array}{ll}
\phi (x) &=  \phi^{in}(x) - \int d^4y \Delta_R(x-y) \cdot K_y\phi(y)\\
&\\
          &=\phi^{out}(x) - \int d^4y \Delta_A(x-y) \cdot K_y\phi(y) ,\\
\end{array}
\eqno{(B.5)} $$ 

\noindent where $K_y= \Box_y -m^2$ denotes the Klein-Gordon operator, and the retarded
and advanced functions $\Delta_R$ and $\Delta_A$, respectively, satisfy the following equations:

$$
\begin{array}{l}
K_x\Delta_R (x) =  K_x\Delta_A (x)=-\delta^4(x) .
\end{array}
\eqno{(B.6)} $$ 

\noindent In the case ii) the mass shell is a branch-cut in the propagator and there are no discrete
or normalizable single particle states.  Instead of a single particle state, for instance,
 we have a superposition
of multi-photon states like a wave packet. Also, as clarified by Bloch and Nordsieck [45], electron
scattering is an inclusive reaction since an electron is accompanied by an infinite number of infrared photons
which escape detection. In this way we have only a limited class of observable quantities in the infared singular
theories.

Now we turn to the reduction formula in the case i). The general idea is to replace a propagator
by a single particle wave function. For instance,

$$
\begin{array}{l}
\langle 0\vert T\left[\phi (x)\phi^{\dagger} (y)\right]\vert 0\rangle \rightarrow \langle s\vert\phi^{\dagger} (y)\vert 0\rangle .
\end{array}
\eqno{(B.7)} $$ 

\noindent We shall call this operation ${\cal R}_s$ and it is realized by 

$$
\begin{array}{ll}
{\cal R}_s\langle 0\vert T\left[\phi (x)\phi^{\dagger} (y)\right]\vert 0\rangle 
&=\int d^4x \langle s\vert \phi^{\dagger} (x)\vert 0\rangle (-i)K_x
\langle 0\vert T\left[\phi (x)\phi^{\dagger} (y)\right]\vert 0\rangle \\
&\\
&=\langle s\vert \phi^{\dagger} (y)\vert 0\rangle , 
\end{array}
\eqno{(B.8)} $$ 

\noindent or generally,

$$
\begin{array}{l}
{\cal R}_s\langle 0\vert T\left[\phi (x)AB\cdots \right]\vert 0\rangle 
 =\langle s\vert T\left[AB\cdots \right]\vert 0\rangle . 
\end{array}
\eqno{(B.9)} $$ 

\noindent When $T$ is replaced by the antichronological symbol $\tilde{T}$, $(-i)K_x$
should be replaced by $iK_x$.

Amplitudes are not what we can directly observe; so we shall study their absolute squares
and expectation values. The reduction formula for the former should realize the following
replacement:

$$
\begin{array}{l}
\langle 0\vert \tilde{T}\left[\phi (u)\phi^{\dagger} (v)\right]\vert 0\rangle 
\langle 0\vert T\left[\phi (x)\phi^{\dagger} (y)\right]\vert 0\rangle \\
\\
\rightarrow \sum_s \langle 0\vert \phi (u)\vert s\rangle \langle s\vert \phi^{\dagger} (y)\vert 0\rangle \\
\\
 \rightarrow \int_{m^2}^{(m+\lambda )^2} dM^2 \rho(M^2) i\Delta^{+}(u-y,M^2) ,\\
\end{array}
\eqno{(B.10)} $$ 

\noindent where $\lambda$ depends on the experimental condition and corresponds to the Bloch-Nordsieck
states in the case 2). In the case 1) the reduction formula is given, for $\lambda<\mu$, by

$$
\begin{array}{l}
\sum_s \int d^4xd^4y \langle 0\vert  \tilde{T}\left[\dots \phi^{\dagger} (y)\right]\vert 0\rangle 
iK_y^{{}^{{}^{\hspace{-.3cm}\leftarrow}}}  \langle 0 \vert\phi (y)\vert s\rangle \\
\\
\times \langle s\vert \phi^{\dagger} (x)\vert 0\rangle (-i)K_x 
\langle 0\vert T\left[\phi (x) \dots \right]\vert 0\rangle \\
\\
= \sum_s \langle 0\vert \tilde{T}\left[\dots \right]\vert s \rangle 
\langle s\vert T\left[\dots \right]\vert 0\rangle  .\\
\end{array}
\eqno{(B.11)} $$ 

\noindent In this case the result does not depend on $\lambda$ and we may take the limit
 $\lambda\rightarrow 0$.

By the same token the reduction formula for expectation values is given, typically, by

$$
\begin{array}{l}
\int d^4xd^4y \langle s\vert \phi^{\dagger} (x)\vert 0\rangle
(-i)K_y^{{}^{{}^{\hspace{-.3cm}\rightarrow}}} \langle 0
\vert T\left[\phi (y)J(z) \phi^{\dagger} (y)\right]\vert 0\rangle 
\times (-i)K_x^{{}^{{}^{\hspace{-.3cm}\leftarrow}}}  \langle 0\vert \phi (x)\vert s\rangle \\
\\
= \langle s\vert J(z)\vert s\rangle .
\end{array}
\eqno{(B.12)} $$

\noindent In order to proceed to infrared singular theories we shall discuss QED in what follows.

We introduce the electron propagator $S_F(p)$ and put

$$
\begin{array}{l}
(i \gamma\cdot p +m)  S_F(p)= a(p^2) + \frac{i\gamma \cdot p}{m} b(p^2) . 
\end{array}
\eqno{(B.13)} $$ 

\noindent Then for the free electron field we have $a=1$ and $b=0$. The behavior of the propagator near
the mass shell is given by

$$
\begin{array}{l}
a(p^2),\ b(p^2)\propto (1+\frac{p^2}{m^2})^c , ~\mbox{with}~ c= -(3-\alpha)\frac{e^2}{8\pi^2} . 
\end{array}
\eqno{(B.14)} $$ 

\noindent This result was obtained by solving Ovsianikov's equation in RG [46]. It shows that
the mass shell $p^2+m^2=0$ is a branch-cut.  Now define

$$
\begin{array}{ll}
D(\partial)= -i \gamma\cdot \partial -m , &  
\tilde{D}(\partial)= i \gamma\cdot \partial -m  .
\end{array}
\eqno{(B.15)} $$ 

\noindent In the absence of the branch-cut we have

$$
\begin{array}{l}
\int d^4xd^4y \langle s\vert \overline{\psi} (x)\vert 0\rangle (-i) D(\partial_x) 
\langle 0\vert T\left[\psi (x) J(z)  \overline{\psi} (y)\right]\vert 0\rangle\\
\\
\times (-i)\tilde{D}(\partial^{{}^{{}^{\hspace{-.25cm}\leftarrow}}}_y)  \langle 0\vert \psi (y)\vert s\rangle \\
\\
= \langle s\vert J(z)\vert s\rangle .
\end{array}
\eqno{(B.16)} $$ 

\noindent In special cases we know the expectation values $\langle J\rangle$ a priori. For instance, the electric charge of the wave-packet-like state $\vert s\rangle$
is known to be equal to $-e$, and $J$ should be replaced by the space integral of the time
component of $J_{\mu}$,

$$
\begin{array}{l}
Q= \int d^3z J_0 (z) , 
\end{array}
\eqno{(B.17)} $$ 

\noindent where $J_{\mu}$ represents the 4-dimensional electric current. In this case $\langle Q\rangle$
is independent of the parameter $\lambda$ indicating the cancellation of the branch-cut between the propagator
and the vertex function provided that they are related to one another through the Ward identity (3.47).

Since we have two electron propagators in the Feynman diagram corresponding to $\langle \psi J_{\mu} \overline{\psi}\rangle$,
we have to eliminate one of them by multiplying $\cal{D}(\partial)$ instead of $D$, where

$$
\begin{array}{l}
{\cal D}(ip) S_F(p)= -1  . 
\end{array}
\eqno{(B.18)} $$ 

\noindent By taking the limit $\lambda\rightarrow 0$ we have

$$
\begin{array}{l}
\int d^4xd^4y \langle s\vert \overline{\psi} (x)\vert 0\rangle (-i){\cal D}(\partial_x) 
\langle 0\vert T\left[\psi (x) J_{\mu}(z)  \overline{\psi} (y)\right]\vert 0\rangle\\
\\
\times (-i)\tilde{D}(\partial^{{}^{{}^{\hspace{-.25cm}\leftarrow}}}_y)  \langle 0\vert \psi (y)\vert s\rangle \\
\\
= \langle s\vert J_{\mu}(z)\vert s\rangle C .
\end{array}
\eqno{(B.19)} $$ 

\noindent Since the normalization of the wave-packet-like state does not guarantee the proper
normalization of the "on-shell-limit state", we have introduced the constant $C$. It has to be determined by
utilizing the known value of $\langle Q\rangle$. In QED we have 

$$
\begin{array}{l}
C= 1+ (3-\alpha)\frac{e^2}{8\pi^2} .
\end{array}
\eqno{(B.20)} $$ 

\noindent In the absence of the asymptotic fields we can apply the same procedure to QCD and obtain the result (3.48)
from the identity (3.14) provided that $M_{\mu}$ represents the spin zero projection
of $\delta\overline{\delta}A_{\mu}$ [27,28] and $\vert\gamma\rangle$ the on-shell-limit of a single quark
or a single gluon state in the sense of Bloch and Nordsieck.


\end{document}